\documentclass[10pt,aps,pra,twocolumn,superscriptaddress,floatfix,showpacs,longbibliography]{revtex4-1}
\usepackage{graphicx}
\usepackage{amsfonts}
\usepackage{amssymb}
\usepackage{amsmath}
\usepackage{txfonts}
\usepackage{lipsum}
\usepackage{color}
\usepackage{wasysym}
\usepackage{hyperref}
\usepackage{bbold}
\usepackage{etoolbox} % for additional refs as footnotes
\graphicspath{{figures/}}

\usepackage{mathtools}%enhanced math capabilities
\usepackage{multirow}%for the tables
\usepackage{hhline}

\newcommand{\av}[1]{\ensuremath{\left\langle #1 \right\rangle}}
\newcommand{\qv}{\mathbf{q}}
\newcommand{\kv}{\mathbf{k}}

\usepackage{letltxmacro}
\LetLtxMacro{\oldsqrt}{\sqrt}
\renewcommand{\sqrt}[2][\mkern8mu]{\mkern-6mu\mathop{}\oldsqrt[#1]{#2}}

\begin{document}

\title{Dual Boson Diagrammatic Monte Carlo Approach applied to the Extended Hubbard model}

\author{M. Vandelli}
\affiliation{The Hamburg Centre for Ultrafast Imaging, Luruper Chaussee 149, 22761 Hamburg, Germany}
\affiliation{I. Institute of Theoretical Physics, Department of Physics, University of Hamburg, Jungiusstrasse 9, 20355 Hamburg, Germany}
\affiliation{Max Planck Institute for the Structure and Dynamics of Matter, Center for Free Electron Laser Science, 22761 Hamburg, Germany}

\author{V. Harkov}
\affiliation{I. Institute of Theoretical Physics, Department of Physics, University of Hamburg, Jungiusstrasse 9, 20355 Hamburg, Germany}
\affiliation{European X-Ray Free-Electron Laser Facility, Holzkoppel 4, 22869 Schenefeld, Germany}

\author{E. A. Stepanov}
\affiliation{I. Institute of Theoretical Physics, Department of Physics, University of Hamburg, Jungiusstrasse 9, 20355 Hamburg, Germany}

\author{J. Gukelberger}
\affiliation{Microsoft Quantum, One Microsoft Way Redmond, WA 98052, USA}

\author{E. Kozik}
\affiliation{Department of Physics, King’s College London, Strand, London WC2R 2LS, UK}

\author{A. Rubio}
\affiliation{Max Planck Institute for the Structure and Dynamics of Matter, Center for Free Electron Laser Science, 22761 Hamburg, Germany}
\affiliation{Center for Computational Quantum Physics, Flatiron Institute, 162 5th Avenue, New York, NY 10010, USA}
\affiliation{Nano-Bio Spectroscopy Group and ETSF, Universidad del Pa\'is Vasco, 20018 San Sebast\'ian, Spain}

\author{A. I. Lichtenstein}
\affiliation{I. Institute of Theoretical Physics, Department of Physics, University of Hamburg, Jungiusstrasse 9, 20355 Hamburg, Germany}
\affiliation{European X-Ray Free-Electron Laser Facility, Holzkoppel 4, 22869 Schenefeld, Germany}
\affiliation{The Hamburg Centre for Ultrafast Imaging, Luruper Chaussee 149, 22761 Hamburg, Germany}

\begin{abstract}
In this work we introduce the Dual Boson Diagrammatic Monte Carlo technique for strongly interacting electronic systems.
This method combines the strength of dynamical mean-filed theory for non-perturbative description of local correlations with the systematic account of non-local corrections in the dual boson theory by the diagrammatic Monte Carlo approach.  
It allows us to get a numerically exact solution of the dual boson theory at the two-particle local vertex level for the extended Hubbard model.
We show that it can be efficiently applied to description of single particle observables in a wide range of interaction strengths. We compare our exact results for the self-energy with the ladder dual boson approach and determine a physical regime, where description of collective electronic effects requires more accurate consideration beyond the ladder approximation. 
Additionally, we find that the order-by-order analysis of the perturbative diagrammatic series for the single-particle Green's function allows to estimate the transition point to the charge density wave phase.
\end{abstract}

\maketitle

\section{Introduction}

Strongly correlated systems represent a formidable challenge in condensed matter physics. 
For this reason, the study of model systems can allow us to investigate the effects of strong interactions and analyse the effects of different approximations. 
Among these models, the Hubbard model~\cite{hubbard} has been extensively studied in the past decades due to its capacity of successfully describing the emerging physics of some classes of strongly correlated materials, where local interactions are assumed to be much stronger than non-local ones. 
A major breakthrough in solution of the Hubbard model was made by dynamical mean-field theory (DMFT)~\cite{RevModPhys.68.13}. This method becomes exact in the limit of infinite spacial dimensions or connectivity of the lattice~\cite{PhysRevLett.62.324}, and serves as an accurate approximation for single-particle quantities in finite dimensions~\cite{PhysRevB.91.235114}.  

At the same time, many real materials exhibit interesting physical effects, such as a charge density wave (CDW) phase, that can not be described by a local Hubbard interaction term alone. In order to consider these phenomena, non-local interactions have to be taken into account. 
For this aim, in analogy with DMFT, an extended dynamical mean field theory (EDMFT) has been developed~\cite{PhysRevB.52.10295, PhysRevLett.77.3391, PhysRevB.61.5184, PhysRevLett.84.3678, PhysRevB.63.115110}. However, in this approach the self-energy and polarization operator are local, meaning that they are frequency dependent, but not momentum dependent.
Extensions of EDMFT, such as the $GW$+EDMFT~\cite{PhysRevB.66.085120,PhysRevLett.90.086402, PhysRevLett.109.226401, PhysRevB.87.125149, PhysRevB.90.195114, PhysRevB.94.201106, PhysRevB.95.245130}, the dual boson (DB)~\cite{RUBTSOV20121320, PhysRevB.90.235135, PhysRevB.93.045107, PhysRevB.100.165128}, the triply irreducible local expansion (TRILEX)~\cite{PhysRevB.92.115109, PhysRevB.93.235124, PhysRevLett.119.166401}, and the dual TRILEX (D-TRILEX)~\cite{PhysRevB.100.205115} approaches have been developed to cope with this issue. In particular, the DB and D-TRILEX techniques are based on the exact transformation that allows to rewrite the initial action of the extended Hubbard model in terms of the local impurity problem, which can be solved numerically exactly~\cite{PhysRevB.72.035122, PhysRevLett.97.076405, PhysRevLett.104.146401, RevModPhys.83.349, HAFERMANN20131280}, and a diagrammatic series around the impurity. Since the effective impurity problem already includes the main contribution coming from local correlations, this looks a naturally convenient starting point for perturbation theory and approximate approaches. 

So far, calculations in the framework of the dual boson approach have been performed only 
in the ladder approximation~\cite{PhysRevLett.113.246407, PhysRevB.90.235105,  PhysRevB.94.205110, NatureVanLoon, NatureZhenya}. This approach is based on the calculation of a specific sub-set of diagrams that, in principle, can be justified by physical considerations only in the regime of developed collective electronic fluctuations~\cite{PhysRevLett.121.037204, PhysRevB.99.115124}. However, the ladder DB approximation provides remarkably good results in comparison with other advanced methods, as for instance dynamical cluster approximation~\cite{PhysRevB.95.115149, PhysRevB.97.115117}. An alternative approach that involves solution of parquet equations based on dual theories was recently proposed in Ref.~\onlinecite{PhysRevB.101.075109}.
Comparison between various methods based on extensions of (E)DMFT can be found in Ref.~\onlinecite{RevModPhys.90.025003}.

Another route to study strongly correlated systems has recently been attempted: the use of unbiased methods based on the combination of diagrammatic approaches with Markov chain Monte Carlo~\cite{PhysRevLett.81.2514}. In particular, the bare diagrammatic Monte Carlo (DiagMC) method has been successfully applied to the Hubbard model at weak and moderate Coulomb interactions~\cite{Kozik_2010}. This method starts from an expansion in terms of the Hubbard coupling $U$ and constructs all Feynman diagrams up to some finite but high order in $U$. The algorithm allows to sample all possible diagrams without any restriction to specific topologies.
Efficient algorithms that express all connected diagrams of the perturbative expansion up to a given order by means of determinants~\cite{PhysRevLett.119.045701} have been developed for various observables and correlation functions ~\cite{PhysRevB.100.121102,PhysRevB.97.085117,sigmaDet,PhysRevLett.124.117602}, significantly reducing the computational cost of the calculation.
Approaches based on a small-coupling expansion work very well in the regime of small to moderate couplings, but start to fail when $U$ is of the order of half of the bandwidth~\cite{PhysRevX.5.041041, PhysRevB.91.235114}. These failure is related to the finite convergence radius of the diagrammatic series and can be improved using resummation techniques~\cite{PhysRevB.100.121102}.

To allow for a non-perturbative treatment of strong correlation effects, a diagrammatic Monte Carlo scheme based on dual fermion (DF) approach~\cite{PhysRevB.77.033101} was proposed in Refs.~\onlinecite{PhysRevB.94.035102, PhysRevB.96.035152}. The advantage of this method in comparison with diagrammatic expansions in terms of the bare Coulomb interaction $U$ is that the impurity problem already accounts for the main effects of local correlations, which strongly screen the bare interaction $U$. The expansion is thus performed in terms of the renormalized local interaction vertex function, which appears to be naturally more convenient at moderate and strongly interacting regime.
Additionally, the diagrams are sampled in continuously in the momentum space without the discretization of the Brillouin zone. Hence, the result of the calculation is not influenced by any finite-size effects.

In this paper we generalize this approach to the extended Hubbard model, performing an additional dual transformation that introduces effective bosonic fields related to non-local interactions.
Our Dual Boson Diagrammatic Monte Carlo (DiagMC@DB) method combines the advantages of DMFT, because it already accounts for the screened local interaction in the impurity problem, with the capability of sampling all the possible Feynman diagrams without any restriction. 
The result is an efficient diagrammatic Monte Carlo algorithm that naturally incorporates non-local Coulomb interaction in the original DiagMC@DF approach~\cite{PhysRevB.96.035152}.

\section{Dual boson theory}

Our starting point is the extended Hubbard model in the action formalism 
\begin{align}
\mathcal{S} =& -\sum\limits_{\kv,\nu,\sigma}c^{*}_{\kv\nu\sigma}\left[i\nu + \mu -\varepsilon_{\kv} \right] c^{\phantom{s}}_{\kv\nu\sigma} \notag\\
&+ U \sum_{\qv,\omega}n_{-\qv,-\omega,\uparrow} \, n_{\qv\omega\downarrow}  +\frac{1}{2}\sum_{\qv,\omega,\varsigma} V^{\varsigma}_{\qv}\, \rho^{\varsigma}_{-\qv,-\omega} \, \rho^{\varsigma}_{\qv\omega}.
\end{align}
Here, $c^{(*)}_{\kv \nu \sigma}$ are Grassman variables corresponding to the annihilation (creation) of electrons with momentum $\kv$, fermionic Matsubara frequency $\nu$ and spin $\sigma$. We also introduced the electronic dispersion $\varepsilon_{\kv}$ and chemical potential $\mu$. 
The model additionally includes an on-site Coulomb interaction of strength $U$ in terms of the electron density $n_{\qv\omega\sigma}$ at momentum $\qv$ and bosonic Matsubara frequency $\omega$,  as well as a non-local interaction $V^\varsigma_\qv$, where the index $\varsigma$ represents charge ($\varsigma={\rm ch}$) or spin ($\varsigma={\rm sp}=\{x, y, z\}$) degrees of freedom. Variables $\rho^\varsigma_{\qv\omega}=n_{\qv\omega}^{\varsigma}-\left\langle n_{\phantom{}}^{\varsigma}\right\rangle$ are expressed in terms of composite quantities $n_{\qv\omega}^{\varsigma} = \sum_{\kv\nu,\sigma\sigma'} c^{*}_{\kv+\qv,\nu+\omega,\sigma} \sigma^{\varsigma}_{\sigma\sigma'} c^{\phantom{*}}_{\qv,\omega,\sigma'}$. In the previous expression $\sigma^{{\rm ch}}=\mathbb{1}$, and $\sigma^{x,y,z}$ is the corresponding Pauli matrix in spin space.
The general idea of dual theories is to split the action into two parts: a local impurity problem, that contains the full local interaction, and a non-local part that can be treated perturbatively. Instead of directly applying a perturbation theory to the non-local part, a transformation that introduces new variables is performed. This allows to dress the non-local part with the local impurity quantities. An additional important consideration is that, once the DMFT impurity is chosen, the dual theories represent a diagrammatic expansion around the DMFT solution. This starting point for the perturbation theory looks appealing, since the DMFT already accounts for local many-body effects, which allows to correctly reproduce both the small and large $U$ limits. 
In order to perform this transformation, we add and subtract an arbitrary fermionic
hybridization function $\Delta_{\nu}$, so that we can isolate a local impurity part of the action.
With this fermionic hybridization function the action reads ${\mathcal{S} = \sum_{i}\mathcal{S}^{(i)}_{\text{imp}}+\mathcal{S}^{\phantom{i}}_{\text{nonloc}}}$, where the impurity part is
\begin{align}
\mathcal{S}_{\text{imp}}^{(i)} = 
-\sum\limits_{\nu,\sigma} c^{*}_{\nu\sigma}\left[i\nu + \mu - \Delta^{\phantom{*}}_{\nu} \right] c^{\phantom{s}}_{\nu\sigma} + 
U \sum_{\omega}n_{-\omega,\uparrow} \, n_{\omega\downarrow},
\label{impurityaction}
\end{align}
and the non-local part reads
\begin{align}
\mathcal{S}_{\rm nonloc} =& -\sum\limits_{\kv,\nu,\sigma}c^{*}_{\kv\nu\sigma}\left[\Delta_\nu -\varepsilon_{\kv} \right] c^{\phantom{s}}_{\kv\nu\sigma} +\frac{1}{2}\sum_{\qv,\omega,\varsigma} V^{\varsigma}_{\qv}\, \rho^{\varsigma}_{-\qv,-\omega} \, \rho^{\varsigma}_{\qv\omega}.
\end{align}
In the following, $\av{\ldots}_{\rm imp}$ denotes the average with respect to the local action~\eqref{impurityaction}.
The impurity problem of Eq.~\eqref{impurityaction} can be solved exactly using continuous-time quantum Monte
Carlo solvers~\cite{PhysRevB.72.035122, PhysRevLett.97.076405, PhysRevLett.104.146401, RevModPhys.83.349, HAFERMANN20131280}.
In the same way we could include a bosonic hybridization function~\cite{RUBTSOV20121320,PhysRevB.90.235135, PhysRevB.93.045107}. However, this step would require an additional discussion of the self-consistency condition needed to determine it. Therefore, we exclude the bosonic hybridization from the current discussion in order to reduce the number of external parameters in the system.
The hybridization function can be defined in an arbitrary way, but some choices are more convenient than others. 
In the rest of the paper we will use $\Delta_\nu$ obtained from single-site DMFT impurity problem.

The dual boson transformation amounts to perform a fermionic and a bosonic Hubbard-Stratonovich transformations over the non-local part of the action ${\cal S}_{\rm nonloc}$, which introduce new dual fermionic variables $f$, $f^*$ and bosonic $\phi^{\varsigma}$ fields. The action obtained after this transformation is quadratic in the electronic operators $c^{(*)}_{}$, so we can integrate them out~\cite{PhysRevB.94.205110}. The original problem of interacting electrons is then recast into a new problem in terms of the dual degrees of freedom only. Sigle- and two-particle observables of the original electron system can be exactly related to dual correlation functions, as shown for example in Ref.~\onlinecite{PhysRevB.94.205110}. The result for the dual action reads (see Ref.~\onlinecite{PhysRevB.100.205115} for the derivation)
\begin{align}
\tilde{\mathcal{S}} = &-\sum\limits_{\kv,\nu,\sigma} f^{*}_{\kv\nu\sigma} \tilde{\mathcal{G}}^{-1}_{ \kv\nu\sigma} f^{\phantom{s}}_{\kv\nu\sigma} - \frac{1}{2}\sum_{\qv,\omega,\varsigma} \phi^{\varsigma}_{-\qv,-\omega} \tilde{\mathcal{W}}_{\qv\omega}^{\varsigma \; -1} \phi^{\varsigma}_{\qv\omega} + \tilde{\mathcal{F}}[f,\phi].
\label{eq:Sdual}
\end{align}
The bare dual propagators are defined as 
\begin{align}
\tilde{\mathcal{G}}^{\phantom{E}}_{\kv\nu\sigma} &
=\left[g_{\nu}^{-1}+\Delta_\nu-\epsilon_\kv\right]^{-1}-g_{\nu}
= G^{\text{EDMFT}}_{\kv\nu\sigma} - g^{\phantom{E}}_{\nu\sigma} , 
\label{eq:G_tild}\\
\tilde{\mathcal{W}}^{\varsigma}_{\qv\omega} &
= \alpha_{\omega}^{\varsigma}\;\left[V_\qv^{\varsigma \; -1}-\chi_{\omega}^{\varsigma}\right]^{-1}  \alpha_{\omega}^{\varsigma}
= W^{\varsigma\, \text{EDMFT}}_{\qv\omega}-{w^{\varsigma}_{\omega}}, 
\end{align}
where $g_\nu$ and $w^{\varsigma}_\omega$ are the Green's function and renormalized interaction of the auxiliary impurity problem, respectively, and the impurity susceptibility $\chi_\omega^{\varsigma} = -\av{\rho^{\varsigma}_{-\omega} \, \rho^{\varsigma}_{\omega}}_{\rm imp}$. Additionally, $\alpha^{\varsigma}_{\omega} = 1+U^\varsigma \, \chi_\omega^{\varsigma}$ with $U^{\rm ch/sp}=\pm U/2$.
The choice of the Matsubara frequency space is natural in this case, because in Eq.~\eqref{eq:G_tild} the $\sim\nu^{-1}$ part of the tail in $G^{\text{EDMFT}}_{\kv\nu}$ is exactly canceled by $g_\nu$. This means that the dual fermion Green's function decays as fast as $\sim \nu^{-2}$, and there are no convergency issues related to summations over Matsubara frequencies.   
The interaction term truncated at the two-particle level is given by
\begin{align}
\label{dualinteraction}
&\tilde{\mathcal{F}}[f,\phi] = \sum_{\substack{\qv,\kv,\omega,\nu\\\varsigma,\sigma,\sigma'}} \Lambda^{\varsigma}_{\nu\omega} f^*_{\kv\nu\sigma} \sigma^{\varsigma}_{\sigma\sigma'} f^{\phantom{*}}_{\kv+\qv,\nu+\omega,\sigma'} \phi^{\varsigma}_{\qv\omega} \notag\\
&+ {\frac{1}{4}} \sum_{\substack{\qv,\omega\\ \{\kv,\nu,\sigma\}}}\Gamma^{\sigma\sigma'\sigma''\sigma'''}_{\nu\nu'\omega} f^*_{\kv\nu\sigma}  f^{\phantom{*}}_{\kv+\qv,\nu+\omega,\sigma'} f^*_{\kv'+\qv,\nu'+\omega,\sigma'''}  f^{\phantom{*}}_{\kv'\nu'\sigma''},
\end{align}
where $\Lambda^{\varsigma}_{\nu\omega}$ and $\Gamma^{\sigma\sigma'\sigma''\sigma'''}_{\nu\nu'\omega}$ are the impurity fermion-boson and fermion-fermion vertex functions, respectively. These quantities are defined in the particle-hole form as in Ref.~\onlinecite{PhysRevB.100.205115}, that in terms of impurity variables explicitly read
\begin{align}
    \Lambda^{\varsigma}_{\nu\omega} &= \frac{\av{c^{\phantom{s}}_{\nu \uparrow} c^*_{\nu+\omega,\uparrow}\, \rho_{-\omega}^{\varsigma}}_{\rm imp}}{g_{\nu} g_{\nu+\omega} \alpha_\omega^{\varsigma}},\\
\Gamma^{\sigma\sigma'\sigma''\sigma'''}_{\nu\nu'\omega} &=   \frac{\av{c^{\phantom{s}}_{\nu \sigma}  c^*_{\nu+\omega,\sigma'} c^*_{\nu'\sigma''} c^{\phantom{s}}_{\nu'+\omega,\sigma'''}}_{\rm c,imp}}{g_{\nu} g_{\nu+\omega} g_{\nu'} g_{\nu'+\omega} }.
\label{impurity4vertex}
\end{align}

In general, the interaction term also contains all the higher order vertices that conserve the number of dual fermions, but we will limit our study to the two-particle interaction terms only. Terms beyond this approximation were shown to lead to very small corrections in many regimes~\cite{PhysRevLett.102.206401}. As a matter of fact, dual theories with only two-particle vertex functions show a rather good agreement with other unbiased methods, and it is still under debate if deviations with other methods are due to higher-order vertices or to different effects~\cite{PhysRevB.96.035152,  PhysRevB.94.035102, PhysRevLett.102.206401,  PhysRevB.97.125114, PhysRevB.96.235127}.
Additionally, the inclusion of higher order vertices would enormously increase the complexity of the diagrammatic Monte Carlo scheme. In light of all these considerations, we exclude them from our calculations.

In our case, the solution of the impurity problem is obtained using a continuous-time Monte Carlo solver based on hybridization expansion (CT-HYB)~\cite{HAFERMANN20131280}. 
This gives an access to all the impurity observables needed for the construction of the dual boson diagrammatics. In particular, we compute $\Delta_\nu$, $g_\nu$, and $\chi_\omega$ for the construction of bare propagators, as well as the fermion-fermion vertex $\Gamma^{\sigma\sigma'\sigma''\sigma'''}_{\nu\nu'\omega}$ and the fermion-boson vertex $\Lambda^{\varsigma}_{\nu\omega}$.
Within this approximation, the dual action~\eqref{eq:Sdual} with the interaction~\eqref{dualinteraction} is quadratic in the bosonic fields.
This means, that it is possible to integrate dual bosonic degrees of freedom out exactly and obtain a fully fermionic action. The bosonic Hubbard-Stratonovich transformation is necessary for decoupling of the non-local interaction term, that otherwise would prevent the integration of the local impurity action out.
Moreover, the introduction of the bosonic variables dresses the interaction in terms of the impurity quantities, so that the bosonic propagator is already partially screened. In order to construct a form of the full fermion-fermion vertex after the integration of the bosons, it is useful to decompose the impurity fermion-fermion vertex~\eqref{impurity4vertex} in channel representation as 
\begin{align}
&{\Gamma}^{\sigma\sigma'\sigma''\sigma'''} = 
{\frac{1}{2}} \sum_{\varsigma}{\Gamma}^{\varsigma} \;  \sigma_{\sigma\sigma'}^\varsigma  \sigma_{\sigma''\sigma'''}^\varsigma.
\label{spinvertices}
\end{align}
\begin{figure}[t!]
\centering
\includegraphics[scale=0.8]{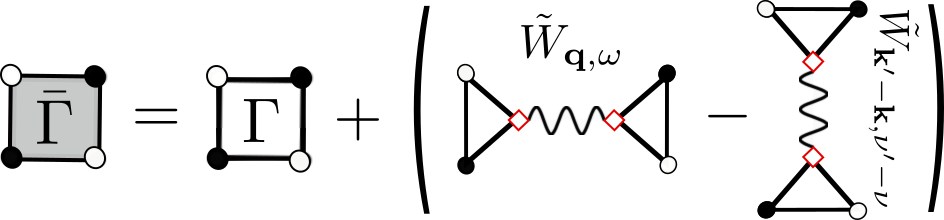}
\caption{\label{fig:vertex} Schematic diagrammatic interpretation of Eq.~\ref{DBvertex}. The full antisymmetrized fermion-fermion interaction $\overline{\Gamma}$ (gray box) is a combination of the impurity vertex ${\Gamma}$ (white box) and processes involving a boson exchange (wiggly line). The full vertex acquires a momentum dependence due to the presence of the bosonic lines. White and black dots represent incoming and outgoing particles, respectively. Triangles represent $\Lambda_{\nu\omega}$ vertices. The exact dependence on the channel indices and prefactors is shown in Eq.~\ref{DBvertex}.}
\end{figure}
The result is a modified dual fermion action 
\begin{align}
&\tilde{\mathcal{S}} = -\sum\limits_{\kv,\nu,\sigma} f^{*}_{\kv\nu\sigma} \tilde{\mathcal{G}}^{-1}_{ \kv\nu\sigma} f^{\phantom{s}}_{\kv\nu\sigma} \notag\\
&+ {\frac{\xi}{8}} \sum_{\substack{\qv,\omega,\varsigma\\ \{\kv,\nu,\sigma\}}}\overline{\Gamma}^{\varsigma, \kv\kv'\qv}_{\nu\nu'\omega} f^*_{\kv\nu\sigma} \sigma_{\sigma\sigma'}^\varsigma f^{\phantom{*}}_{\kv+\qv,\nu+\omega,\sigma'} f^*_{\kv'+\qv,\nu'+\omega,\sigma'''} \sigma_{\sigma'''\sigma''}^\varsigma f^{\phantom{*}}_{\kv'\nu'\sigma''},
\label{finalaction}
\end{align}
where $\xi$ is a formal expansion parameter, which must be set to unity ($\xi=1$) in the actual calculations and keeps track of the expansion order. Importantly, we introduced a new momentum dependent fermion-fermion vertex that combines the vertex function of the local impurity problem and the non-local interaction between fermions mediated by dual bosonic fields
\begin{align}
\overline{\Gamma}^{{\rm ch}, \kv\kv'\qv}_{\nu\nu'\omega} &= \Gamma^{\rm ch}_{\nu\nu'\omega} + 2 \tilde{M}_{\nu,\nu',\omega}^{{\rm ch},\qv} - \tilde{M}_{\nu,\nu+\omega,\nu'-\nu}^{{\rm ch}, \kv'-\kv} - 3\tilde{M}_{\nu,\nu+\omega,\nu'-\nu}^{{\rm sp}, \kv'-\kv}, \notag\\
\overline{\Gamma}^{{\rm sp}, \kv\kv'\qv}_{\nu\nu'\omega} &= \Gamma^{{\rm sp}}_{\nu\nu'\omega} + 2\tilde{M}_{\nu,\nu',\omega}^{{\rm sp},\qv} + \tilde{M}_{\nu,\nu+\omega,\nu'-\nu}^{{\rm sp}, \kv'-\kv} - \tilde{M}_{\nu,\nu+\omega,\nu'-\nu}^{{\rm ch}, \kv'-\kv}.
\label{DBvertex}
\end{align}
Here, $\tilde{M}_{\nu,\nu',\omega}^{\varsigma, \qv} = \Lambda^{\varsigma}_{\nu,\omega} \tilde{W}^{\varsigma}_{\qv\omega}\Lambda^{\varsigma}_{\nu'+\omega, -\omega}$.
Since Eq.~\eqref{spinvertices} holds for both vertices $\Gamma_{\nu\nu'\omega}$ and $\overline{\Gamma}^{\,\kv\kv'\qv}_{\nu\nu'\omega}$, it is possible to switch easily between the two representations using the relations
\begin{align}
\Gamma^{\rm ch} = {\Gamma}^{\uparrow\uparrow\uparrow\uparrow} +{\Gamma}^{\uparrow\uparrow\downarrow\downarrow}, ~~ \Gamma^{\rm sp} = {\Gamma}^{\uparrow\uparrow\uparrow\uparrow} - {\Gamma}^{\uparrow\uparrow\downarrow\downarrow}.
\end{align}
Additionally, all the other non-zero components can be simply obtained by applying the ${\rm SU}(2)$ symmetry in Eq.~\eqref{spinvertices}
\begin{align}
{\Gamma}^{\uparrow\downarrow\uparrow\downarrow} =  {\Gamma}^{\uparrow\uparrow\uparrow\uparrow}-{\Gamma}^{\uparrow\uparrow\downarrow\downarrow} = \Gamma^{\rm sp}
\end{align}
or by exploiting the fact that a simultaneous flipping of all the spins leads to the same result in paramagnetic case.
We note that the structure of the new fermion-fermion vertex function~\eqref{DBvertex} shown in Fig.~\ref{fig:vertex} is reminiscent of the D-TRILEX theory~\cite{PhysRevB.100.205115}, and appears due to the antisymmetrized form of the interaction.
Integrating out of bosonic fields is very important for the calculation of diagrams, because it allows to eliminate the bosonic degrees of freedom from the theory analytically and to avoid their sampling in diagrammatic Monte Carlo.

In our implementation we compute the dual self-energy. In order to obtain the single-particle observables of the lattice problem, we can use the standard equation that relates the dual self-energy to the lattice self-energy $\Sigma_{\kv\nu}$,
\begin{align}
   \Sigma_{\kv\nu} = \Sigma_\nu^{{\rm imp}}+\overline{\Sigma}_{\kv\nu}
 \label{latticeselfenergy}
\end{align}
with $\overline{\Sigma}_{\kv\nu} = \frac{ \tilde{\Sigma}_{\kv\nu}}{1+g_\nu\,\tilde{\Sigma}_{\kv\nu} }$,  where $\Sigma_\nu^{{\rm imp}}$ is the self-energy of the impurity problem, as shown for example in Ref.~\onlinecite{PhysRevB.94.205110}. The lattice Green's function can be obtained via the usual Dyson equation from the lattice self-energy or using its relation with the dual Green's function~\cite{PhysRevB.94.205110}.

\section{Diagrammatic Monte Carlo scheme \label{DiagMCscheme}}

The algorithm tested in this paper is an extension of the DiagMC@DF method proposed in Refs.~\onlinecite{PhysRevB.96.035152, PhysRevB.94.035102}. Our DiagMC algorithm computes numerically exactly the coefficients $a_n(\mathbf{k}, \nu)$ in the expansion of the dual self-energy
\begin{align}
\tilde{\Sigma}_{\mathbf{k} \nu}(\xi) = \sum_n a_n(\mathbf{k}, \nu) \, \xi^n, \label{eq:sigma_dual_series}
\end{align} 
for the action~\eqref{finalaction} up to some maximum order $N_{max}$. The value of the dual self-energy can be recovered by setting $\xi=1$.
We will call it diagrammatic Monte Carlo for dual bosons (DiagMC@DB). In the same way as the original algorithm, our method is based on bare diagrammatic Monte Carlo approach~\cite{PhysRevB.77.020408}. This algorithm allows to construct all the Feynman diagrams up to any finite order and to sum over them using Markov chain Monte Carlo. According to Refs.~\onlinecite{PhysRevLett.81.2514, FedorSimkovic}, any correlation function $\mathcal{O}$ can be expressed as a sum of diagrams as follows 
\begin{align}
\mathcal{O}(y) &= \lim\limits_{N_{max} \rightarrow +\infty} \sum_{n=0}^{N_{max}}\sum_{\lbrace x_i \rbrace_{}} \mathcal{O}_{\mathcal{C}_n}(\lbrace x_i \rbrace, y) = \notag\\ &=\lim\limits_{N_{max} \rightarrow +\infty}\sum_{n=0}^{N_{max}}\sum_{\lbrace x_i \rbrace_{}} \text{sgn}\left(\mathcal{O}_{\mathcal{C}_n}(\lbrace x_i \rbrace, y)\right) \,\cdot \,\left|\, \mathcal{O}_{\mathcal{C}_n}(\lbrace x_i \rbrace, y)\, \right|,
\label{observables}
\end{align}
where $y$ is a combined index that contains all the dependence on external points, $n$ indicates the number of vertices that appear in the diagram, $\mathcal{C}_n$ are the topologies, and $\mathcal{O}_{\mathcal{C}_n}$ is the value of a specific diagram. Additionally $x_i$ is shorthand notation for the internal degrees of freedom $(\kv, \nu, \sigma)_i$ corresponding to momentum, Matsubara frequency and spin that originate from the presence of loops of Green's functions.
This statement is true provided that the limit in Eq.~\eqref{observables} is well defined and convergent for the chosen parameters as $N_{max} \rightarrow +\infty$. Divergencies of the diagrammatic series are often related to physical instabilities, as we show in Sec.~\ref{sec:CDW}, or to some unphysical behavior of the starting point, for example the antiferromagnetic phase transition of DMFT~\cite{PhysRevB.96.035152}.

The summation over the perturbation order $n$, topologies and internal degrees of freedom is performed using a Metropolis-Hastings scheme~\cite{PhysRevB.77.020408}, where the function to be sampled is the $\text{sgn}\left(\mathcal{O}\right)$, and the probability distribution is given by he amplitude $\left|\, \mathcal{O}\right|$ in order to respect the requirement of positive weight function. This approach automatically satisfies the detailed balance condition for the Markov chain {(see Ref.~\onlinecite{10.1093/biomet/57.1.97})}, given that the acceptance probability to go from a configuration $\mathcal{C}$ to another configuration $\overline{\mathcal{C}}$ is constructed as
\begin{align}
R_{{\mathcal{C} \longrightarrow \overline{\mathcal{C}}}} = \text{min}\left\lbrace \; 1, \; \frac{\mathcal{P}_{\overline{\mathcal{C}}}}{\mathcal{P}_{ \mathcal{C}}}\cdot \frac{|\mathcal{O}_{\overline{\mathcal{C}}}|}{|\mathcal{O}_{\mathcal{C}}|}\right\rbrace,
\end{align}
where $\mathcal{P}_{ \mathcal{C}}$ and $\mathcal{P}_{\overline{\mathcal{C}}}$ are the probabilities of the initial and final configuration respectively. There are no substantial changes from DiagMC@DF in the acceptance-rejection scheme adopted, except for the fact that in our case the bare fermion-fermion vertex function~\eqref{DBvertex} is momentum dependent. Each contribution to the series expansion~\eqref{observables} can be written as a combination of two kinds of diagrammatic elements: fermionic lines that represent dual Green's functions $\tilde{\mathcal{G}}$ (called also propagator lines) and vertices $\overline{\Gamma}$ described in Eq.~\eqref{DBvertex}. Each vertex is attached to four propagator lines, two incoming and two outgoing. The terms at order $n$ in the expansion are represented in terms of Feynman diagrams with $n$ vertices connected by lines in all the possible combinations. 

These diagrams give an intuitive and efficient picture that allows us to design the updates so that all the contributions to the expansion~\eqref{observables} can be generated by changing how the vertices are connected to each other by mean of the propagator lines. 
In particular, we use the same worm algorithm described in the Ref.~\onlinecite{PhysRevB.96.035152} to update the diagram topologies. The aim of the worm algorithm is to enforce momentum conservation, which is a non-local property of the diagram, by means of updates that act locally on few elements of a diagram. The worm algorithm introduces a set of unphysical updates that allow the transition between all the different possible topologies contributing to the dual self-energy $\tilde{\Sigma}_{\kv\nu}$. This means that we sample all the diagrams with one incoming and one outgoing line that are also irreducible with respect to a cut of a fermionic line. This can be practically implemented by the condition that no internal line can carry the same momentum and frequency dependence of the external lines.
All the subtleties and details related to the implementation are discussed in detail in Ref.~\onlinecite{PhysRevB.96.035152}. 

Each configuration is identified by an ordered set of vertices, where each vertex is stored together with the incoming and outgoing frequencies, momenta, spins and the connections with the other vertices. The original implementation of Ref.~\onlinecite{PhysRevB.96.035152} works with unsymmetrized diagrams, in order to avoid topology-dependent prefactors. However, the local vertex $\Gamma$ itself has an antisymmetric form in spin space, and we find convenient to introduce also the non-local vertex corrections $\tilde{M}^{\varsigma,\,\qv}_{\nu,\nu'\omega}$ in the antisymmetrized form shown in Eq.~\eqref{DBvertex}. The corresponding vertex in the unsymmetrized diagrammatic theory can be obtained by simply dividing this vertex by two~\cite{PhysRevB.96.035152}.

\begin{figure}[t!]
\centering
\includegraphics[width=0.75\linewidth]{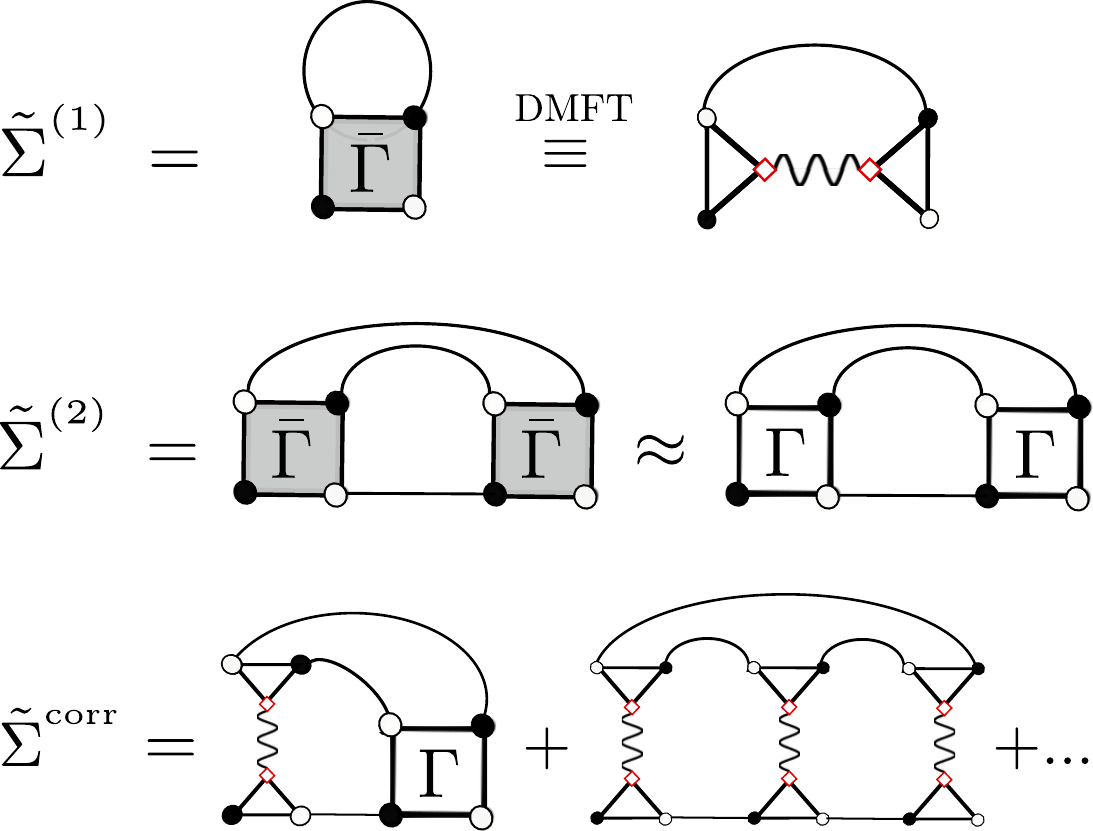}
\caption{\label{fig:diagrams} Most important self-energy diagrams. Top row shows the only nonzero contribution to the first order diagram, taking into account that we can not connect two slots of the same local vertex with a propagator line due to DMFT self-consistency condition. The middle row shows the second-order diagrams $\tilde{\Sigma}^{(2)}$. If $V$ is small compared to $U/4$, it can be approximated by the second-order dual fermion diagram on the right hand side. The last term $\tilde{\Sigma}^{ {\rm corr}}$ shows few diagrams that enter $\tilde{\Sigma}$ in our calculations, but are not included in the ladder DB.}
\end{figure}

The simultaneous sampling of contributions coming from the fermion-fermion scattering and boson exchange processes efficiently reduces the number of topologies.
On the other hand, we do not distinguish between local and non-local diagrams, so that we can not exclude local diagrams from the beginning simply by a proper choice of the DMFT self-consistency condition, as it was done in DiagMC@DF calculations. 
However, in the spirit of Refs.~\onlinecite{VANHOUCKE201095, Kozik_2010}, we can reduce the diagrammatic space and thereby increase the efficiency of DiagMC sampling by self-consistently eliminating all diagrams that contain insertions of the topology $\Sigma^{(1)}$ depicted in Fig.~\ref{fig:diagrams}. This is accomplished by the so-called ``semi-bold'' DiagMC scheme of Ref.~\onlinecite{PhysRevB.93.161102}, in which the bare Green's function in all diagrams is replaced with $\mathcal{G}_{\rm{sb}}$ that is dressed by the first-order self-energy, found as the self-consistent solution of the Dyson equation $\tilde{\mathcal{G}}^{-1}_{\rm{sb}} = \tilde{\mathcal{G}_0}^{-1} - \Sigma^{(1)}[\tilde{\mathcal{G}}_{\rm{sb}}] $. Here $\Sigma^{(1)}[\tilde{\mathcal{G}}_{\rm{sb}}]$ is the first order diagram where the bare propagator is replaced by the self-consistently calculated one $\tilde{\mathcal{G}}_{\rm{sb}}$. This formal transformation of the series is exact in the sense that it does not change the final result~\cite{PhysRevB.93.161102}, although the convergence properties of the semi-bold series are generally different~\cite{PhysRevB.96.041105}.
\begin{figure}[t!]
\flushleft
\includegraphics[width=1\linewidth, trim={1cm 2cm 1.5cm 2cm}]{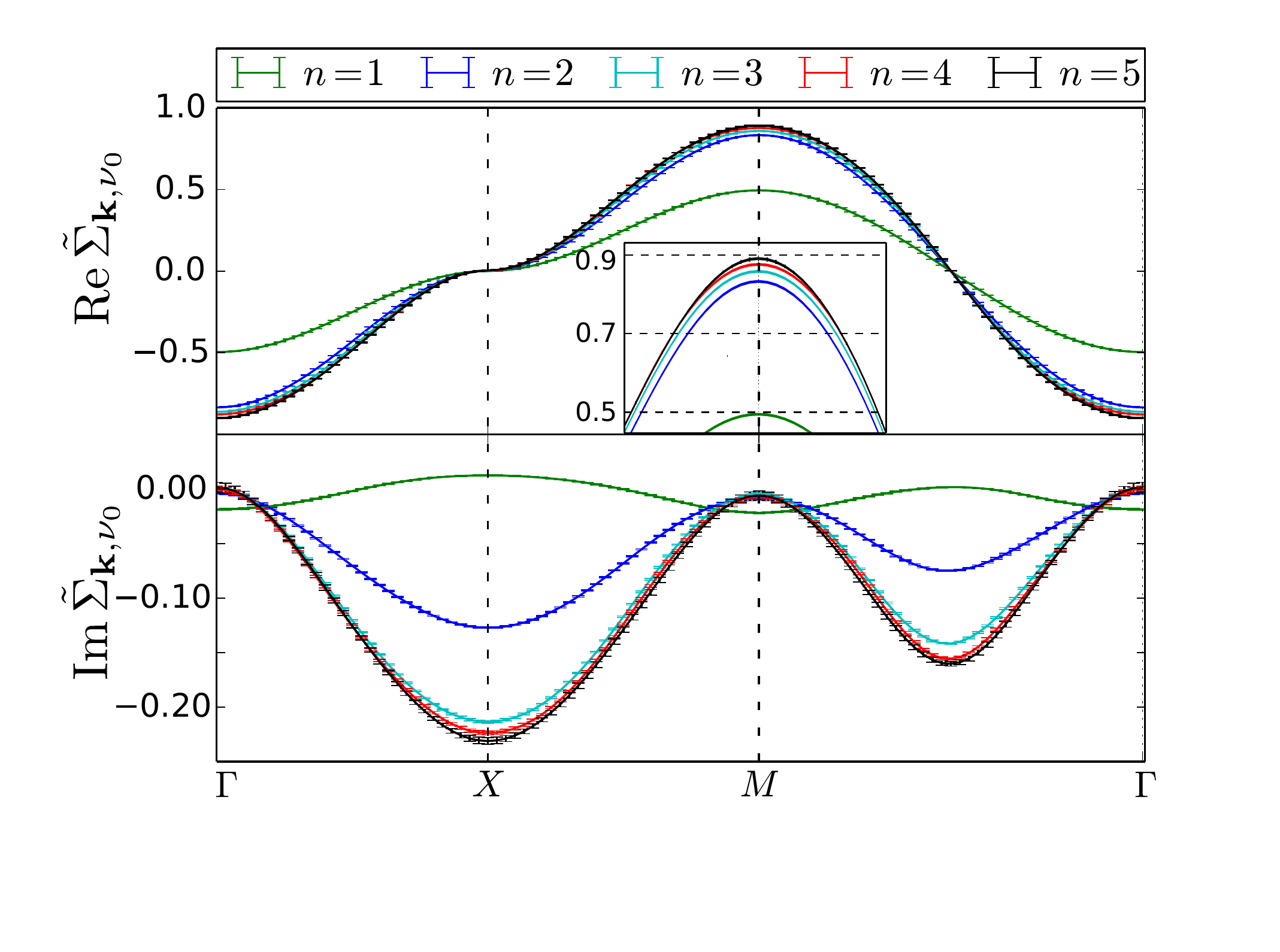}    
\caption{\label{dualsigma} Convergence of the real (top panel) and imaginary (bottom panel) parts of the dual self-energy $\tilde{\Sigma}_{\kv,\nu}$ obtained for the zeroth Matsubara frequency $\nu_0$. The result is plotted along the high-symmetry path in momentum space $\kv$ as a function of the expansion order $n$. The parameters are $U=5$, $V=1.25$, and $\beta=2$ in the units of the hopping amplitude. The inset shows the convergence of the real part around the $M=\{\pi, \pi\}$ point. }
\end{figure}
Using the Metropolis-Hastings scheme allows us to compute observables up to a normalization factor. In order to keep track of the normalization, we sample the absolute value of an additional diagram that we can calculate explicitly outside Monte Carlo and we store its value in a suitable accumulator $N_{\rm norm}$.
The chosen diagram is simply a single vertex with unitary value with the upper corners connected by a single bare dual Green's function.
Its value is given by
\begin{align}
\mathcal{N} = \sum_{\kv\nu}\left\vert\tilde{\mathcal{G}}_{\kv\nu}\right|,
\end{align}
which is computed directly from the analytical expression for the bare dual propagator $\tilde{\mathcal{G}}$. 
The normalized dual self-energy $\tilde{\Sigma}_{\kv \nu}$ is then straightforwardly computed from the normalization accumulator $N_{norm}$ using the following equation 
\begin{align}
\tilde{\Sigma}_{\kv \nu} = \frac{\mathcal{N}}{N_{\rm norm}} \left\langle \tilde{\Sigma}_{\kv \nu} \right\rangle_{\text{MC}}.
\end{align}
 
\section{Results \label{sec:CDW}}
\subsection{Computational details}

We perform our calculations on a 2D square lattice with the nearest-neighbor hopping amplitude $t=1$ that fixes the energy units. The chemical potential $\mu$ is set to $U/2$, ensuring that the system is at half-filling. In order to avoid the low temperature issues related to the DMFT N\'eel transition discussed in details in Ref.~\onlinecite{PhysRevB.96.035152}, all the calculations are performed at $\beta=4$ when $U \leq 4$ and at $\beta=2$ for $U>4$. 
We would like to stress, that this is not a limitation of the method, which works with any dispersion and with a general form of the interaction as a function of momentum. 
We start from the description of the output of the calculation, namely the dual self-energy $\tilde{\Sigma}_{\kv\nu}$ obtained within the bare diagrammatic Monte Carlo scheme. The only obvious difference between a bare and and semi-bold run is that the $\Sigma^{(1)}[\tilde{\mathcal{G}}_{\rm{sb}}]$ is computed in advance and added to the DiagMC result. Since the latter sums diagrams up to a given order, one has to check that the result is converged with respect to the order.
In Fig.~\ref{dualsigma} we show a converged output of our calculations. In particular, the result for the maximum order of the diagrammatic expansion, order 5, differs from order 4 of $\sim1\%$, hence we consider the result converged. Practically, this means that the performed calculation can be considered converged already at order 4. This is the case for most of presented calculations. 

Far away from instabilities, it is not possible to observe any improvement in going beyond the 5th order of the diagrammatic expansion, and the computation necessary to achieve convergence at the order 6 increases significantly. Thus, a standard DiagMC@DB computation requires around 12 hours with a hundred parallel runs in order to obtain a converged result at the 5th order. Instead, for a converged result at the 6th order, the required computational time increases to more than 24 hours in order to reach a reasonable accuracy. For this reason, all results presented in this work are calculated up to 5th order of expansion, except the ones that are used for the analysis of the phase transition. 

\begin{figure}[t!]
\centering   
\includegraphics[width=0.9\linewidth]{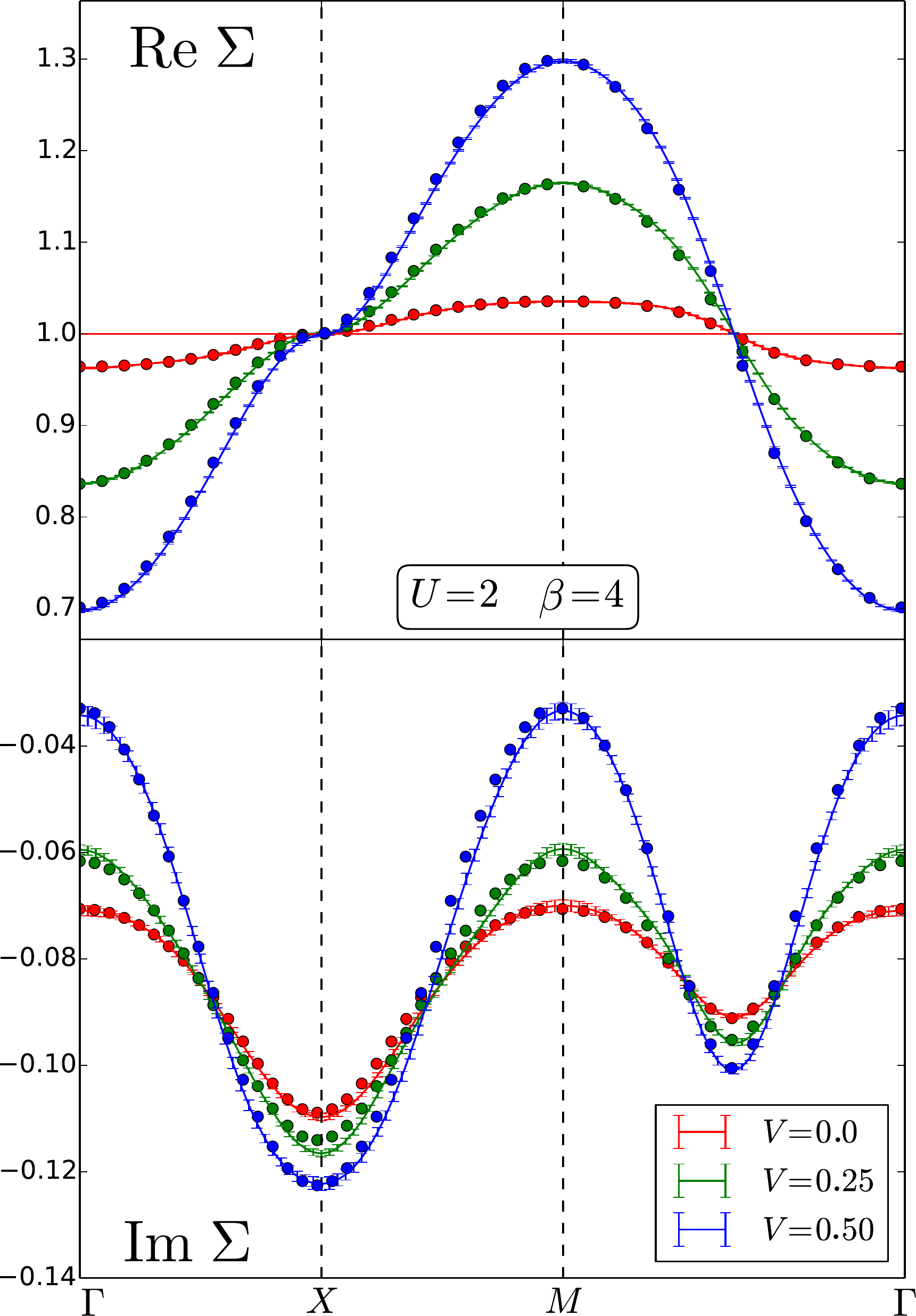}
\caption{\label{fig:comparison2} Comparison between DiagMC@DB (solid lines with error bars) and ladder DB (dots) results for the real (top panel) and imaginary (bottom panel) parts of the lattice self-energy $\Sigma_{\kv \nu_0}$. The result is obtained for $U=2$, $\beta=4$, and different values of the non-local Coulomb interaction $V$ specified in the legend of the bottom panel.}
\end{figure}

\begin{figure}[t!]
\centering   
\includegraphics[width=0.875\linewidth]{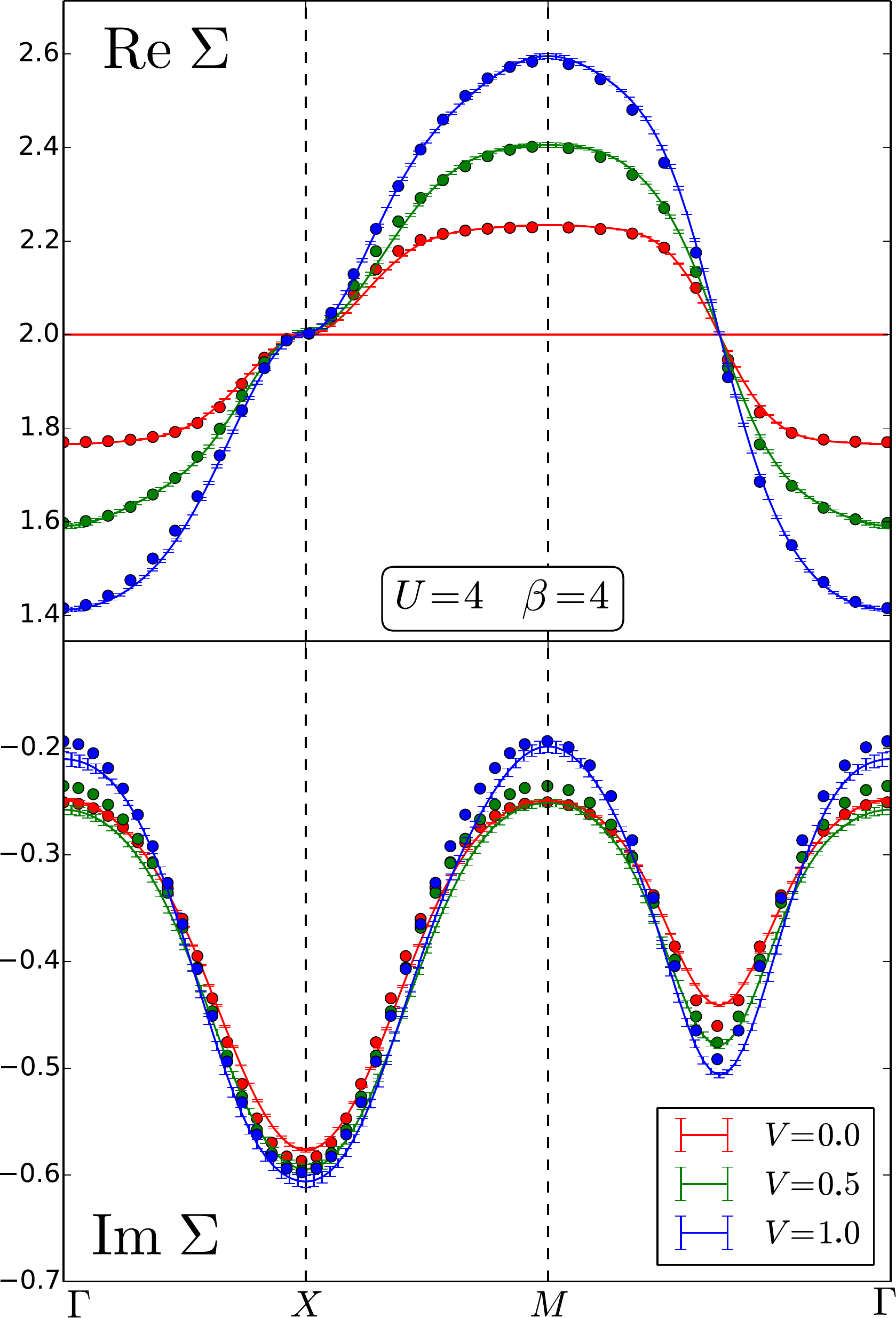}  
\caption{\label{fig:comparison4} Comparison between DiagMC@DB (solid lines with errorbars) and ladder DB (dots) results for the real (top panel) and imaginary (bottom panel) parts of the lattice self-energy $\Sigma_{\kv \nu_0}$. The result is obtained for $U=4$, $\beta=4$, and different values of the non-local Coulomb interaction $V$ specified in the legend of the bottom panel.}
\end{figure}

An important remark is that the contribution coming from the non-local interaction can be quite large, even up to values around $U=6t$. 
Additionally, we observe that the main contribution to the real part of the self-energy comes from two kind of diagrams that are shown in Fig.~\ref{fig:diagrams}. 
The first is the single boson diagram $\tilde{\Sigma}^{(1)}$ that contains only one factor $\tilde{M}^{\varsigma,\,\qv}_{\nu,\nu'\omega}$, which is the only non-zero contribution at the first order of the diagrammatic expansion in terms of the vertex function.
This can be already seen in Fig.~\ref{dualsigma}, where the first order contribution accounts for around 50\% of the real part of the dual self-energy. 
The second important contribution is the second order dual fermion diagram $\tilde{\Sigma}^{(2)}$, that contains two fermion-fermion vertices connected to each other.
At values of $V$ far away from the CDW instability, other contribution to ${\rm Re }\tilde{\Sigma}$ are rather small compared to these ones. 
On the other hand, the imaginary part of the self-energy is much more sensitive to higher order corrections. 
In Fig.~\ref{dualsigma} we can see that the second order is way off compared to the third order, accounting for only around 50\% of the contributions to ${\rm Im}\tilde{\Sigma}$.
Interestingly, the third order already accounts for most of the contributions. 
We deduce, that the inclusions of third-order diagrams in our expansion that contain multiple fermion-fermion scattering and bosonic exchanges are important for the imaginary part of the self-energy. 
These diagrams contribute to around 40\% of ${\rm Im}\tilde{\Sigma}$ at high symmetry points for $U=5$, and their impact on dual quantities becomes even more important at larger $U$. 
Orders larger than the third typically amount to a correction of less than 10\% of ${\rm Im}\tilde{\Sigma}$ at high symmetry points.

However, the overall momentum dependence of the lattice self-energy is still dominated by ${\rm Re}\tilde{\Sigma}$ in the regimes where $U \leq 4$ or $U \geq 8$, because of the denominator in Eq.~\eqref{latticeselfenergy}, as shown also in Ref.~\onlinecite{PhysRevB.96.035152}. 
The most important corrections to the lattice self-energy coming from ${\rm Im}\tilde{\Sigma}$ appear exactly in the regime between half of the bandwidth and the bandwidth, where it can account for around 40\% of the difference with DMFT solution at high symmetry points. 
Even though second-order is thought to already account for the most important contributions far away from instabilities, as shown in Ref.~\onlinecite{PhysRevB.91.235114}, the inclusion of two-boson exchanges and third-order corrections in fermion-fermion vertices can lead to significant quantitative improvements over second-order calculations.  

\subsection{Comparison with the ladder DB approach}

\begin{figure}[t!]
\centering   
\includegraphics[width=0.9\linewidth]{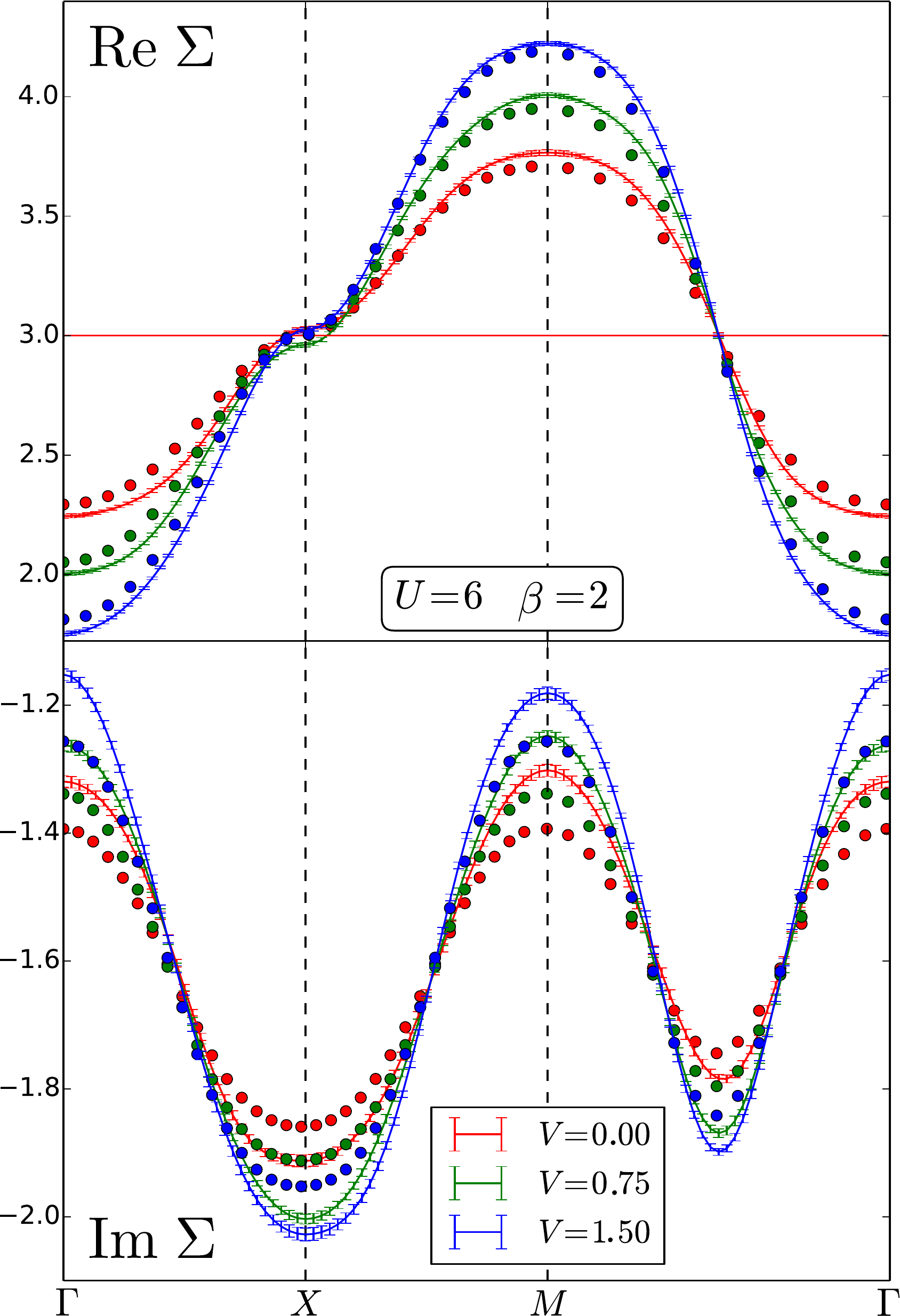} 
\caption{\label{fig:comparison6} Comparison between DiagMC@DB (solid lines with errorbars) and ladder DB (dots) results for the real (top panel) and imaginary (bottom panel) parts of the lattice self-energy $\Sigma_{\kv \nu_0}$. The result is obtained for $U=6$, $\beta=2$, and different values of the non-local Coulomb interaction $V$ specified in the legend of the bottom panel.}
\end{figure}

\begin{figure}[t!]
\centering   
\includegraphics[width=0.9\linewidth]{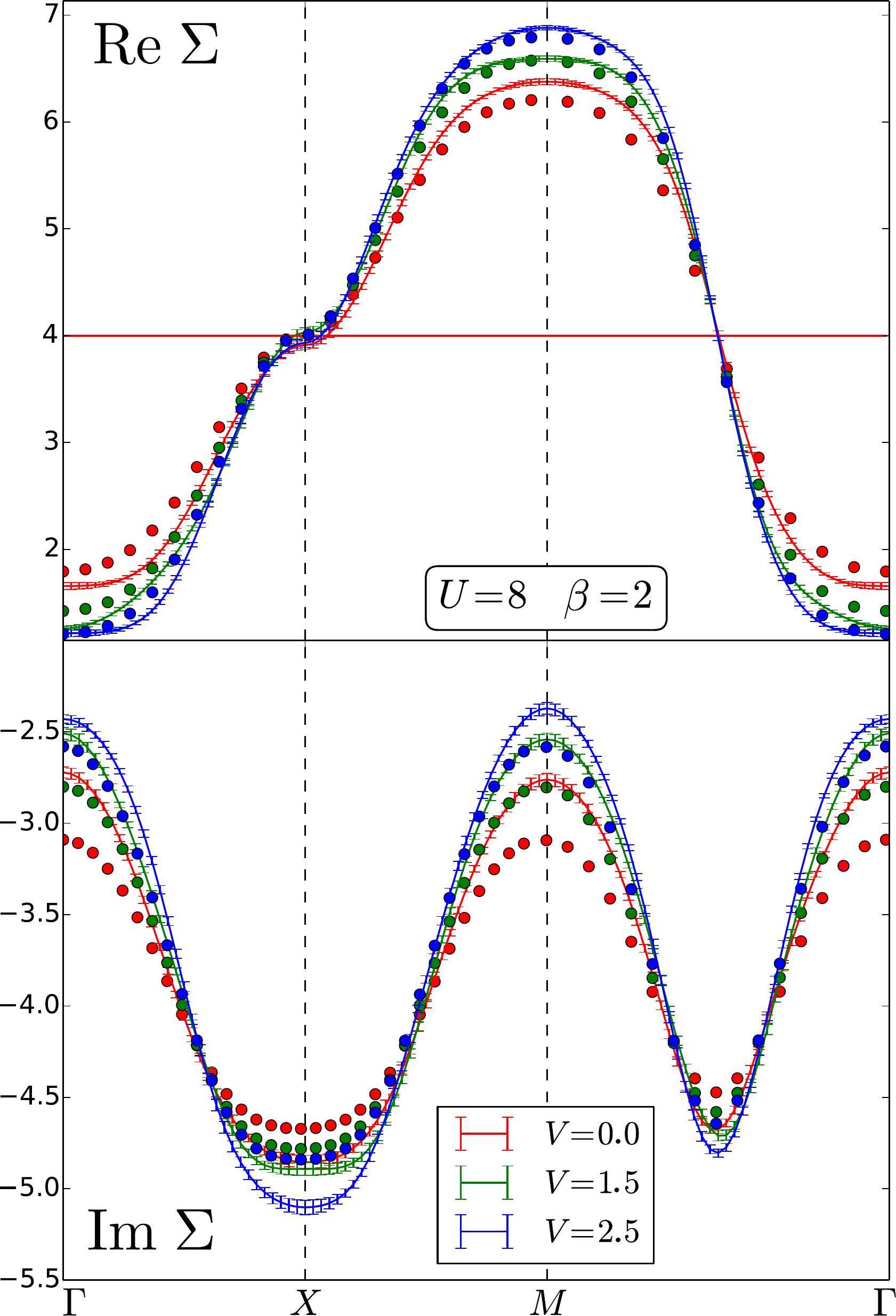} \\
\caption{\label{fig:comparison8} Comparison between DiagMC@DB (solid lines with errorbars) and ladder DB (dots) results for the real (top panel) and imaginary (bottom panel) parts of the lattice self-energy $\Sigma_{\kv \nu_0}$. The result is obtained for $U=8$, $\beta=2$, and different values of the non-local Coulomb interaction $V$ specified in the legend of the bottom panel.}
\end{figure}

Figures~\ref{fig:comparison2}-\ref{fig:comparison8} show a comparison of the DiagMC@DB and ladder DB calculations for different values of the local $U$ and non-local $V^{\rm ch}_{\qv}$ Coulomb interactions. In all the figures we show the result of the calculation at the 5th order. In particular, we show results from a quarter of the bandwidth $U=2$ up to the bandwidth $U=8$. 
We note that the agreement between these two methods is substantially exact up to a half of the bandwidth for all considered values of the non-local interaction. 
In fact, in this regime the ladder DB result for the lattice self-energy lies inside the error bars of the DiagMC@DB calculation. 
For larger values of the on-site Coulomb interaction exceeding the half of the bandwidth, the difference between two theories is more noticeable, especially for small strength of the non-local interaction $V$.
In order to quantify the difference between these two methods, we look at the $M=(\pi,\pi)$ point in the momentum space and calculate the following quantity
\begin{align}
    \delta_{M} ={\rm Re}\left[ \frac{\overline{\Sigma}_{M,\nu_0}^{\rm{DiagMC}}-\overline{\Sigma}_{M,\nu_0}^{\rm{ladd.}\phantom{g}}}{\overline{\Sigma}_{M,\nu_0}^{\rm{DiagMC.}\phantom{g}}}\right],
\end{align}
where $\overline{\Sigma}_{M,\nu_0}$ is the difference between the self-energy in the specified method and the DMFT self-energy $\Sigma_{\nu_0}^{\rm imp}$~\eqref{latticeselfenergy}, and $\nu_0$ is the lowest positive Matsubara frequency.
We measure differences with respect to DMFT self-energy, because the latter is constant in momentum space and quite large. If we want to resolve relatively small differences in momentum space, we have to exclude its contribution.
Additionally, we choose the $M$ point, because it shows the largest difference between the two curves in the Brillouin zone. In this way, we are sure that the $\delta_M$ parameter contains information only about the maximum mismatch coming from the dual corrections. The reason for taking the real part of this quantity comes from the fact that the imaginary part of the dual self-energy ${\rm Im} \tilde{\Sigma}$ shows a systematic shift already at $V=0$, i.e. at the dual fermion level (see Ref.~\onlinecite{PhysRevB.96.035152}).
Here, we aim to assess the behaviour of the self-energy as a function of the non-local $V$ rather than to investigate this aspect.

The result for the mismatch parameter $\delta_{M}$ is summarised in a tentative phase diagram shown in Fig.~\ref{fig:comparison}. 
We can conclude that the difference between DiagMC@DB and ladder DB approaches is negligible at small $U$ below the half of the bandwidth and further increases with the local interaction. 
This behavior can be explained considering that for small local Coulomb interaction $U$ the regime is still perturbative in the dual boson theory, so we expect all the methods to give quantitatively similar results. 
This finding is also in agreement with the result of DiagMC@DF calculations~\cite{PhysRevB.96.035152} obtained for the zero non-local Coulomb interaction. 
On the other hand, we observe that the mismatch is more severe at $V=0$ and decreases as $V$ increases. 
Indeed, when the non-local Coulomb interaction is large, charge fluctuations in the horizontal channel are expected to give the main contribution to physical observables such as self-energy and susceptibility~\cite{PhysRevB.99.115124}, because the system lies close to the charge density wave (CDW) phase. 
Ladder DB approach accounts for this kind of fluctuations by construction, and for this reason the mismatch $\delta_M$ decreases. 
From Fig.~\ref{fig:comparison}, we find that the values of $U$ at which the largest mismatch occurs (red area) lie in the region where the phase transition to the Mott-insulating state was predicted by cluster DMFT ~\cite{PhysRevLett.101.186403} and dual fermion~\cite{PhysRevB.99.205133} calculations at lower temperature. 
This means, that in this regime contributions not included in the ladder approximation cease to be negligible. 
These contributions corresponds to bosonic lines in a direction orthogonal to the ladder direction (see, e.g. $\tilde{\Sigma}^{\rm corr}$ in Fig.~\ref{dualsigma}), which are included in DiagMC@DB. 
It means, that the correct description of this phase transition, especially at $V=0$ would require advanced approaches beyond the ladder approximation.
Another consideration that emerges from this analysis as a function of $U$ and $V$ is that up to a half the bandwidth a momentum dependence of the real part of the self-energy at $V=U/4$ is dominated by the non-local interaction $V$. 
It plays a very important role even for $U=6$, where we would expect the local interaction to give the most important contribution.

\subsection{Monitoring the CDW phase transition from single particle observables }

In the current implementation, the DiagMC@DB theory is based on a solution of a single-site impurity problem, which does not allow for a description of broken-symmetry phases. 
A practical example of the failure of an expansion based on single-site DMFT is given by the strong antiferromagnetic fluctuations arising in the Hubbard model at low temperature, as discussed in detail in Ref.~\onlinecite{PhysRevB.96.035152} for the DiagMC@DF.
In particular, this results in a divergence of the infinite diagrammatic expansion in terms of bare dual quantities at the phase transition~\cite{PhysRevB.90.235135, PhysRevB.94.205110, PhysRevB.99.115124, PhysRevB.100.165128, PhysRevB.96.035152}. 
The most interesting phase of the extended Hubbard model that is associated with the presence of the non-local Coulomb interaction is the charge density wave phase, 
i.e. a checkerboard configuration in the real space with alternating empty sites and doubly occupied sites. 
The phase transition to this state occurs when $V$ is large enough to overcome the effect of the on-site Coulomb repulsion that favours a single-electron occupation of lattice sites. 
A perturbative expansion at small values of $U$ predicts the onset of the CDW phase to be located at $V\simeq{}U/8+\text{const}$~\cite{PhysRevB.99.115112}. A mean-field estimate based on RPA or $GW$ theories gives the transition point at $V\simeq{}U/4$ \cite{PhysRevB.95.245130}. 
This behavior is reproduced at moderate interaction strength by DCA calculations~\cite{PhysRevB.99.245146, PhysRevB.97.115117}. 
Finally, as we shall see below, for large values of $U$ and large temperatures the position of the onset of the CDW phase appears to shift towards the value $V\simeq{}U$ that can be found, for example, using the Peierls-Feynman-Bogoliubov variational principle~\cite{PhysRevLett.111.036601}. Dual boson calculations are in good agreement with the DCA results and reproduce all these different trends in the different regimes~\cite{PhysRevB.90.235135, PhysRevB.94.205110}.

\begin{figure}[t!]
\centering   
\includegraphics[width=0.95\linewidth]{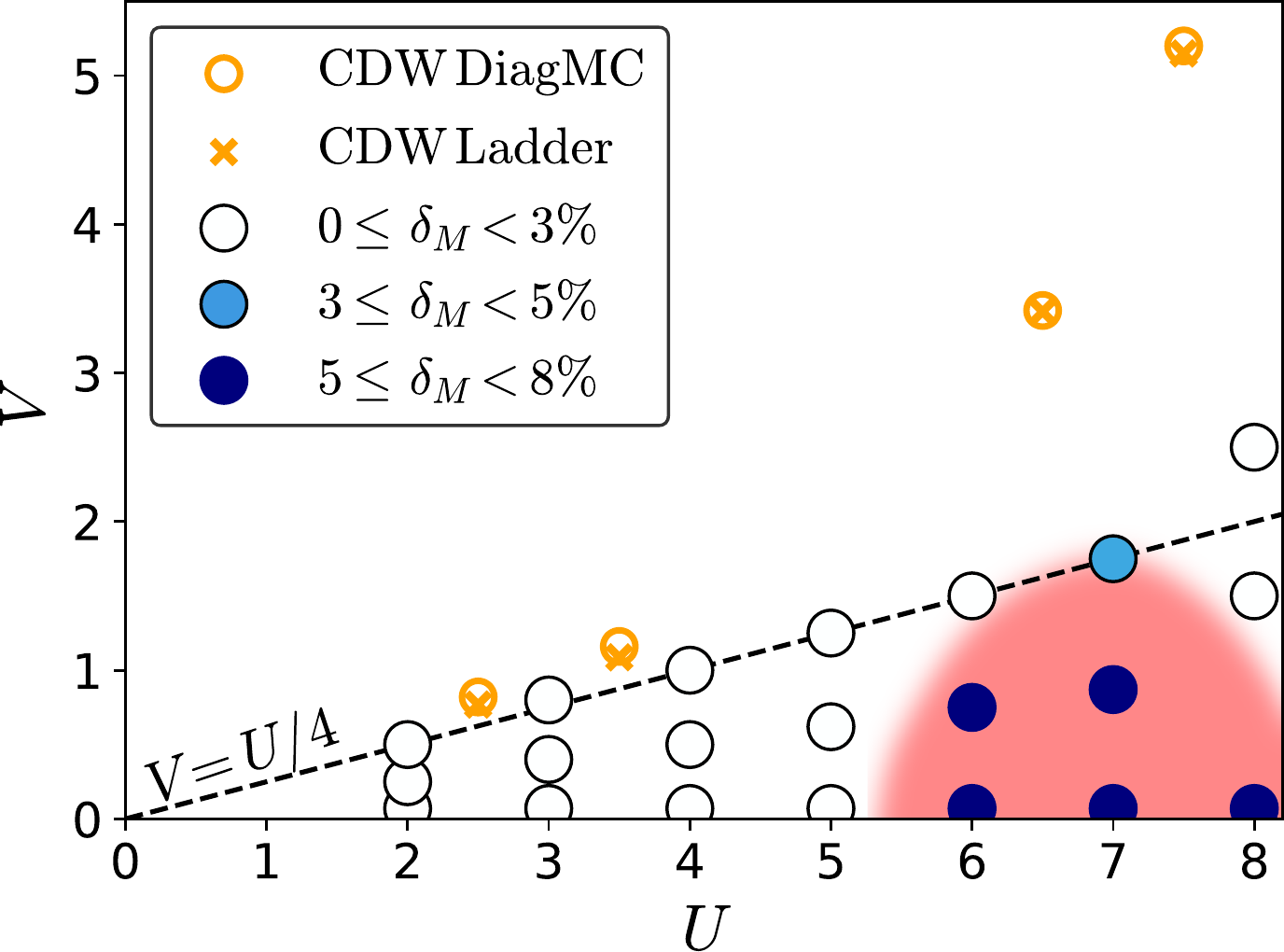} 
\caption{\label{fig:comparison} Summary of the results of our calculations as a function of the local $U$ and non-local $V$ Coulomb interactions. Results for $U \leq 4$ were obtained at $\beta=4$, while for $U>4$ they were calculated at $\beta=2$. The mismatch parameter $\delta_{M}$ is depicted by color. Points correspond to physical parameters for which calculations are performed. The red area highlights the region where the mismatch parameter is larger. Transition points between the normal and CDW phases obtained in DiagMC@DB and ladder DB calculations are depicted by an orange circle and cross, respectively. The dashed black line $V=U/4$ represents an estimation of the phase boundary predicted by mean-field arguments. }
\end{figure}

\begin{figure}[t!]
\includegraphics[width=0.95\linewidth]{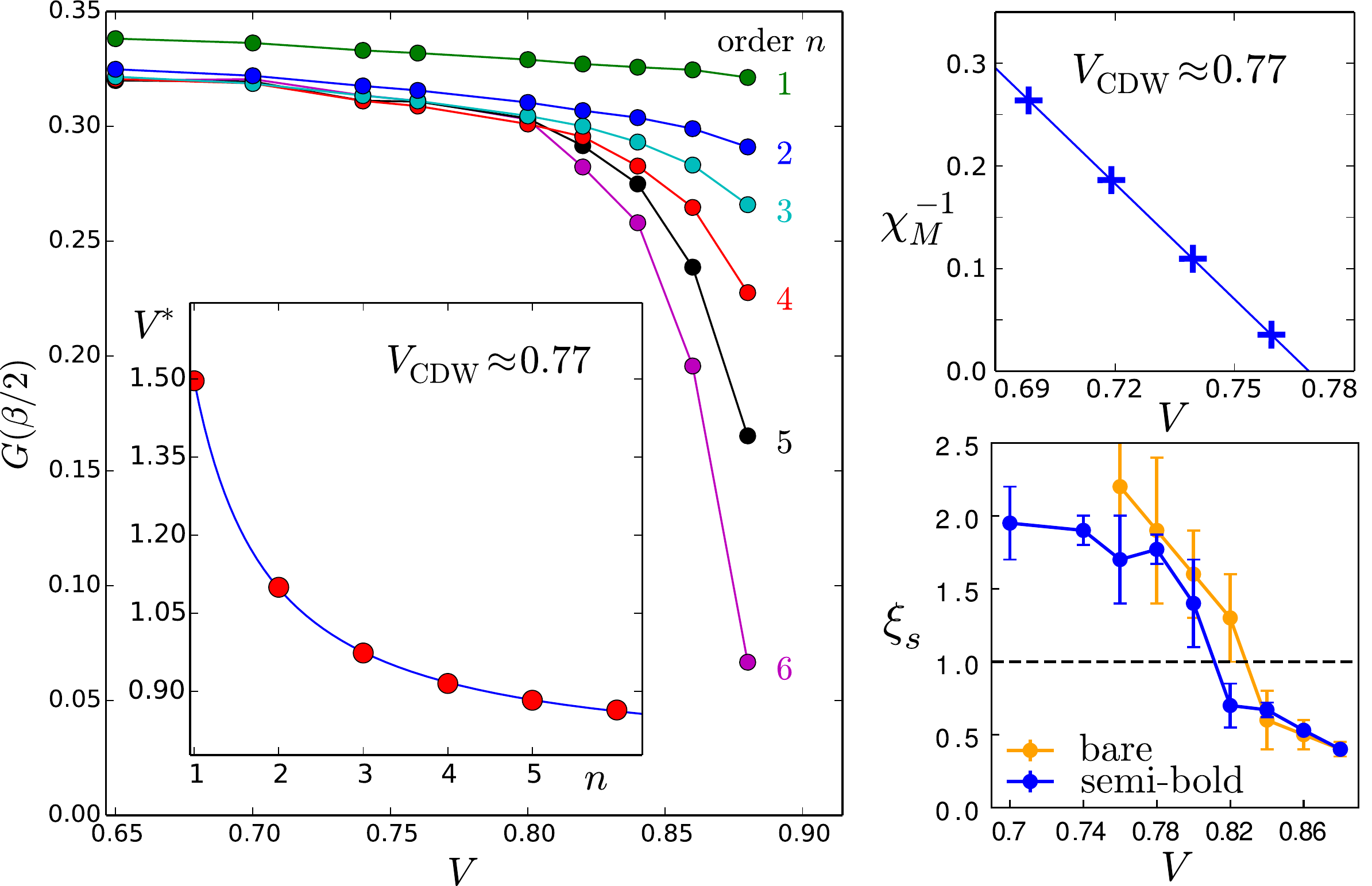}~
\caption{\label{fig:detailed_vs_order} Left panel: $G_{\text{loc}}(\tau=\beta/2)$ as a function of non-local interaction $V$ for the various perturbation orders in the bare DiagMC scheme at $U=2.5$ and $\beta=4$. In the inset it is shown the estimated $V$ of the CDW transition as a function of the order. An additional fitting of the curve with the function $f(n) = C_0 + C_1 \, n^{-C_2}$, where $C_0$, $C_1$ and $C_2$ are fitting parameters, allows to extrapolate the value at infinite order $V_{\rm CDW} =C_0= 0.77(2)$. Upper right panel: The inverse of the charge susceptibility at the $M=\{\pi,\pi\}$ point obtained by ladder dual boson calculations as a function of $V$ for the same $U$ and $\beta$. The value $V_{\rm CDW}$ can be obtained by a linear fitting of the data and checking where the fitting line crosses zero. Lower right panel: Position $\xi_s$ of the singularity on the real axis obtained within the method presented in Ref.~\onlinecite{baker1961Dlog} for the bare and semi-bold DiagMC schemes. The phase transition occurs when $\xi_s$ crosses $\xi = 1$. This analysis predicts the transition at $V_{\rm CDW } = 0.82(1)$ for the bare series and $V_{\rm CDW } = 0.81(1)$ for the semi-bold series.}
\end{figure}

The description of the system inside the CDW state requires an inclusion of symmetry breaking terms in the theory. However, the instability can be identified already in the normal phase studying the charge susceptibility~\cite{PhysRevB.90.235135, PhysRevB.94.205110, PhysRevB.100.165128}. In particular, we expect the susceptibility to show a very sharp peak when the instability occurs. 
This trend can be seen in the upper right panel of Fig.~\ref{fig:detailed_vs_order}, where the inverse of the charge susceptibility linearly decreases. 
Our ladder dual boson calculations predict a transition point $V_{\rm CDW}$ at $V=0.77$ for $U=2.5$ and $V=1.09$ for $U=3.5$ at inverse temperature $\beta=4$. In the strongly correlated regime, the transition points evaluated with this method are $V=3.41$ for $U=6.5$ and $V=5.15$ for $U=7.5$ at $\beta=2$.

However, the critical value $V_\mathrm{CDW}$ for the CDW phase transition can also be found in a controlled way from the analytic structure of the dual self-energy $\tilde{\Sigma}$ as a function of the complex expansion parameter $\xi$. 
Since the dual action~\eqref{finalaction} is explicitly constructed for the translationally-symmetric phase, the critical point $V=V_\mathrm{CDW}$ is marked by a singularity appearing in the function $\tilde{\Sigma}(\xi)$ at $\xi=1$. 
When $V$ is increased beyond $V_\mathrm{CDW}$ in the symmetry-broken phase, this singularity must move along the real axis towards the origin for the physical $\tilde{\Sigma}(\xi=1)$ to remain inaccessible by its power-series expansion (\ref{eq:sigma_dual_series}). 
The method introduced in Ref.~\onlinecite{baker1961Dlog} and routinely applied in the context of DiagMC~\cite{PhysRevB.100.121102} allows to accurately evaluate the specific location of the singularity $\xi_s$. 
It assumes a generic power-law behavior near the singularity, which is typical for a continuous phase transition, $\tilde{\Sigma}(\xi) \propto (\xi_s-\xi)^\eta$ for $|\xi-\xi_s| \ll 1$ with some real number $\eta$, and extracts $\xi_s$ from the behavior of the series coefficients $a_n$ in Eq.~(\ref{eq:sigma_dual_series}). 
As shown in Ref.~\onlinecite{PhysRevB.100.121102}, $\xi_s$ can be found from a finite number of coefficients $\{a_n\}$ with a reliable error bar that includes both the systematic and statistical (Monte Carlo) error.
The result of this procedure for different values of $V$ is shown in the bottom right panel of Fig.~\ref{fig:detailed_vs_order}, where $\xi_s(V)$ is obtained from $\{\mathrm{Re} \, a_n(\mathbf{k}, \nu_0)\}$, for $n=1, \ldots , N_{max}=6$ projected on the first $A_{1\rm g}$-symmetric harmonic $\psi^{s}_{\rm (1,0)}({\bf k}) = \cos(k_x) + \cos(k_y)$ to produce a numerical series from the functional one. 
The condition $\xi_s(V)=1$ then gives the critical value $V_\mathrm{CDW}$. In order to study the behavior of the series close to the phase transition, it is crucial to get very well converged coefficients and to achieve high orders in the expansion. For this reason the calculations of the phase boundary were performed using the semi-bold scheme described in Sec.~\ref{DiagMCscheme} up to the sixth order. The critical values $V_{\rm CDW}$ obtained with this method are $V = 0.81(1)$ at $U=2.5$ and $\beta=4$. In the same fashion we can estimate the transition to occur at $V = 1.15(1)$ for $U=3.5$, $V = 3.42(1)$ for $U=6.5$ and $V = 5.20(2)$ for $U=7.5$. These points are highlighted in orange in Fig.~\ref{fig:comparison}.

In combination with this controlled method, we propose an additional empirical way to obtain the instability point. It is important to notice that the checkerboard configuration of electrons is insulating.
This means that strong charge fluctuations create a pseudogap in the electronic spectrum, which can be detected calculating the spectral function. 
In particular, the spectral function at the Fermi energy is directly connected to the local Green's function $G(\beta/2)$ calculated at imaginary time $\tau=\beta/2$ by the relation $A(E_\text{F})\approx -\beta \, G(\beta/2)/\pi$ (see for example Ref.~\onlinecite{PhysRevB.90.235135}), without the need of analytical continuation from Matsubara to real frequencies.
This situation is conceptually similar to the antiferromagnetic pseudogap, but the analysis in the framework of our theory is very different. 
In fact, the divergence of the dual fermion series is not associated with a true physical instability, as discussed previously, hence the divergence of the diagrammatic series in terms of the local interaction does not have a clear physical interpretation. 
On the contrary, the non-local interaction $V$ enters only the effective fermion-fermion vertex function of Eq.~\eqref{DBvertex} in a trivial form through the dual boson propagator, which up to a local prefactor is proportional to~\cite{PhysRevB.94.205110}
\begin{align}
\tilde{\mathcal{W}}^{\varsigma}_{\qv\omega} \sim \frac{V^{\varsigma}_{\qv}}{1 - \chi^{\varsigma}_{\omega}V^{\varsigma}_{\qv}}. 
\end{align}
We consider the local Green's function obtained by replacing the dual self-energy up to order $n$ into the Dyson equation $G^{(n)}(\beta/2)$ as a function of the non-local interaction $V$, keeping the local interaction $U$ fixed. 
If we inspect left panel of Fig.~\ref{fig:detailed_vs_order}, the behaviour of this function resembles a Fermi function
\begin{align}
 G^{(n)}_{ V}(\beta/2) \approx \frac{G^{(n)}_{V=0}(\beta/2)}{\exp\left[\alpha_n (V-V^*_n)\right]+1}
\end{align}
where $V^*_n$ is the critical value of the non-local interaction at which the function shows a steep drop, and $\alpha_n$ is a numerical value that defines the broadening of the Fermi function at order $n$.
From these empirical and physical considerations, we expect that $V^*_n \rightarrow V_{\rm CDW}$ as the order $n \rightarrow +\infty$, which means that, if we extrapolate the central point of the Fermi function as a function of $n$, we expect it to converge to the value $V_{\rm CDW}$.
Results based on this analysis for the bare series are plotted in the left panel of Fig.~\ref{fig:detailed_vs_order}, and the expected behaviour is clearly visible in the figure.
Fitting the value of $V^*$ as a function of the order $n$ allows to extrapolate the value of $V_{\rm CDW}$ by letting the order go to infinity.
In the case of the semi-bold series, there is a redistribution of weight between the various orders and the extrapolation to infinite order from the first six coefficients based on the same fitting function described in the caption of Fig.~\ref{fig:detailed_vs_order} is not as accurate as in the bare case.
The inset in the left panel of Fig.~\ref{fig:detailed_vs_order} clearly shows the convergence of the values of $V_n^*$ to a finite value as the order increases. 
We compared the extrapolated values from the bare series with the susceptibility at $M=(\pi,\pi)$ calculated from ladder calculations. 
With this simple analysis we obtained values for $V_{\rm CDW}$ compatible with the ladder results. 
The specific values are $V_{\rm CDW} = 0.77(2)$ at $U=2.5$ and $V_{\rm CDW} = 1.09(2)$ at $U=3.5$ with $\beta=4$. 

The methods presented in this section show how order-by-order or cumulative analysis of the series allows for an accurate extrapolation of results in the limit of infinite order of perturbation expansion.
Our results for $U=2.5$ show a very good agreement with the value obtained with $GW$+DMFT in Ref.~\onlinecite{PhysRevB.95.245130}. 
Additionally there is a good agreement with previous dual boson~\cite{PhysRevB.90.235135, PhysRevB.94.205110,PhysRevB.99.115124} and DCA calculations~\cite{PhysRevB.95.115149, PhysRevB.99.245146}. 

\section{Conclusions and outlook}

Even though the method presented in the previous sections is not exact, because higher-order impurity vertices are neglected, the DiagMC@DB scheme allows to include contributions coming from all the possible diagrams with no restriction on a particular class of topologies. 
In other words, DiagMC@DB is the exact solution of the dual boson action truncated at the level of two-particle scattering. 
Due to this consideration, the results provided by DiagMC@DB are based on theoretically much more stable grounds than other approximations based on partial resummation of specific diagrams, as in the case of the ladder dual boson. 
Additionally, there are no finite-size effects since we worked in momentum space and Matsubara frequencies directly. 

In our calculations we observed a very accurate agreement between DiagMC@DB and ladder calculations for $U$ up to half of the bandwidth. 
Even above this value of $U$, the ladder dual boson seems to capture the main contributions and the difference between the two methods is quite small. 
In fact, we have never observed a value of the $\delta_{\rm M}$ parameter bigger than $8\%$ in the region of the parameter space where series converges. 
This offers a further validation of the ladder dual boson technique over a very wide range of interaction strengths. 
The presence of strong non-local interaction $V$ enhances non-local bosonic excitations in the charge channel that are accounted for in ladder approximation. 
This can be captured looking at the real part of the self-energy, which coincides for the two methods at large values of $V$. 

The advantage of the DiagMC@DB is that is allows to consider diagrams order by order and investigate the convergence properties of the series in an unbiased and systematic way. 
In particular, starting from DMFT impurity, the solution is a series is terms of a complicated function of $V$. 
Since $V$ does not enter the impurity, it is possible to use resummation techniques to estimate the value of $V$ at which the charge order occurs already from the study of single-particle quantities. 
Different choices of the hybridization functions, obtained for instance from ladder dual boson calculations, could in principle extend the convergence radius of the series at lower temperatures (see Ref.~\onlinecite{PhysRevB.96.035152}). 

Another strategy that could improve the efficiency of sampling could be the formulation of the series in terms of the semi-bold Green's function, in which some diagrams are included in the bare dual propagator from the very beginning, or the fully bold Green's function, substantially reducing the configuration space. 
It is expected that both approaches could improve the convergence properties as well, but a systematic study is required. In our calculations for the phase diagram, we observe that the computational time needed for a converged result at the sixth order is typically decreased by an order of magnitude if the first order diagram is included in the semi-bold Green's function. At the same time,  this choice of the semi-bold scheme consistently gives results compatible with the bare series in the whole parameter space investigated in this study.

We are currently working in the direction of extending this method to calculate also two-particle observables in two-dimensional heterostructures. 
In addition, the inclusion of a checkerboard configuration with two non-equivalent sublattices (impurity problems) can allow to study the Extended Hubbard model inside the broken symmetry phases, as the CDW phase or the antiferromagnetic phase.

\begin{acknowledgments}
We acknowledge financial support from the European Research Council (ERC-2015-AdG-694097).
This work is funded  by the Cluster of Excellence 'CUI: Advanced Imaging of Matter' of the Deutsche Forschungsgemeinschaft (DFG) - EXC 2056 - project ID 390715994.
Additional support by the Max Planck Institute, the North-German Supercomputing Alliance (HLRN) under the Project No. hhp00042, and the Simons Foundation is also acknowledged.
The Flatiron Institute is a division of the Simons Foundation.
\end{acknowledgments}

\bibliography{ref.bib}

\end{document}